\documentclass[twocolumn,showpacs,preprintnumbers,pra,amsmath,amssymb]{revtex4}

\usepackage{graphicx}
\usepackage{bm}
\begin{document}
\title{Quantum State Discrimination with General Figures of Merit}

\author{Won-Young Hwang$^{1,2}$ \footnote{Email: wyhwang@jnu.ac.kr} and Yeong-Deok Han$^{3}$}

\affiliation{$^{1}$Department of Physics Education, Chonnam National University, Gwangju 500-757, Republic of Korea\\
$^{2}$ Institute for Quantum Computing, University of Waterloo, Waterloo, ON N2L 3G1, Canada\\
$^{3}$ Department of Game Contents, Woosuk University, Wanju, Cheonbuk 565-701, Republic of Korea\\
}
\begin{abstract}
We solve the problem of quantum state discrimination with ``general (symmetric) figures of merit'' for an even number of symmetric quantum bits with use of the no-signaling principle. It turns out that conditional probability has the same form for any figure of merit. Optimal measurement and corresponding conditional probability are the same for any monotonous figure of merit.
\end{abstract}
\maketitle
\section{Introduction}
 The most fundamental element in communications is a task in which for a receiver (Bob)  discriminates among different physical entities sent by a sender (Alice). In general, however,  different quantum states cannot be discriminated with certainty because of non-orthogonality. Interestingly, this fundamental limit makes quantum communication more intriguing and more useful in some cases \cite{Nie00}.

 Let us consider the task.
 If the number of entities is a finite number $N$, the task is normally called quantum state discrimination (QSD) \cite{Hol74, Yue75, Che00}.  If the number of entities is infinite and continuous, it is called quantum state estimation (QSE) \cite{Mas95,Che00}. QSD maximizes the probability of correct guessing for the entity. In QSE, however, the probability of correct guessing is zero because the number of entities is infinite, and thus is a meaningless quantity. Hence a function (a figure of merit) is introduced which assigns a score for the pair of a state sent by Alice and a state guessed by Bob. Normally the figure of merit is monotonous, that is, as the two states are closer, the score is higher. The average score is optimized.

Now let us consider the QSD in the context of a figure of merit.  In the normal QSD, only those events in which Bob makes a correct guess are counted. QSD does not consider how close the guessed one is to the correct one. Here we can generalize the QSD, by introducing a function, the figure of merit, which assigns a score for the pair of a correct state and a guessed one. We maximize the average score as in QSE.
As we see, normal QSD is a special case of ``QSD with general figures of merit''. (``QSD with general figures of merit'' is equivalent to ``QSE with discrete set of states''.)

Recently it has been shown that the no-signaling (no superluminal communication) principle can greatly simplify analysis in quantum information \cite{Gis98,Bar02} including analysis of QSD \cite{Hwa05, Bae08, Hwa10, Bae11} and QSE \cite{Han10}.
In this paper, we consider the problem of the ``QSD with general figures of merit'', for an even number $2M$, of symmetric states of quantum bits (qubits), with use of the no-signaling principle. Here both methods used for QSD \cite{Hwa05, Bae08, Hwa10, Bae11} and QSE \cite{Han10} are combined to get solutions. It turns out that optimal measurements are the same for all monotonous (symmetric) figures of merit in QSD.

It is true that the no-signaling principle is not essential in our argument. That is, ``no-signaling principle'' can be replaced by ``impossibility of discriminating two different decompositions of states corresponding to the same density operator'' in Refs. \cite{Hwa05, Bae08, Hwa10, Bae11, Han10} and throughout this paper. However, we adopt the no-signaling principle here because it makes the result more concrete.

This paper is organized as follows. In Sec. II, we introduce a set of symmetric (mixed) qubit states. We show that with use of the no-signaling principle, conditional probability in QSD for the symmetric set always has the form $\alpha \cos^2 (\phi/2)+ \beta$ for any (symmetric) figure of merit. Here $\alpha,\beta$ are constants and $\phi$ is determined by the Bloch vectors of the prepared qubit and the guessed qubit. We also bound the conditional probability with use of the no-signaling principle again. We observe that the function giving the maximal score is realized by a simple symmetric measurement, which is just the optimal measurement in normal QSD. Thus the measurement becomes an optimal measurement in QSD with any monotonous figure of merit.
\section{main contents}
\subsection{Conditional probability has a unique form.}
Let us give a description of QSD. In this paper, $\{q_i, \rho_i\}$ denotes the situation in which $N$ different quantum states $\rho_i$ are generated with probability $q_i$ by Alice, where $\sum_{i} q_i=1$. $P(j|i)$ is a conditional probability that an output $j$ is given for an input $i$ by certain optimal measurements of Bob. Here $i,j=1,2,...,N$. The guessing probability denotes the maximal probability of correct guessing:
\begin{equation}
P_{\mbox{guess}}= \mbox{max} \sum_{i=1}^{N} q_i P(i|i),
\label{Y}
\end{equation}
where maximization is done over measurements.
In the normal QSD, the guessing probability is maximized (or, equivalently, error probability is minimized).
However, we can generalize QSD by introducing a figure of merit, a function $f(i,j)$. The average score is maximized:
\begin{equation}
 S \equiv \sum_{i=1}^{N} q_i \{\sum_{j=1}^{N} f(i,j) P(j|i)\}.
\label{Z}
\end{equation}
In this paper we consider only symmetric figures of merit. That is, we assume that $f(i,j)=f(j,i)$ and $f(i,j)=f(i+n,j+n)$ (mod $N$) for each $i,j$ and $n=1,2,. . .,N$.
The normal QSD corresponds to a case where $f(i,j)=\delta_{ij}$. Here $\delta_{ij}=1$ when $i=j$ and $\delta_{ij}=0$ when $i\neq j$.

Then let us describe Bloch representation \cite{Nie00} in which any qubit state can be expressed as
\begin{equation}
\rho(\vec{r}) = \frac{1}{2} (\openone +\vec{r} \cdot \vec{\sigma}).
\label{A}
\end{equation}
Here, $\vec{r} =|\vec{r}| (\sin\theta\cos\varphi,\sin\theta\sin\varphi,\cos\theta)$ and $\vec{\sigma}=(\sigma_x, \sigma_y, \sigma_z)$, where $\sigma_x, \sigma_y, \sigma_z$ are Pauli operators. $|\vec{r}|=1$ for pure states and $|\vec{r}|<1$ for mixed states.
An even number $N (=2M)$ of symmetric qubit states to be discriminated can be parameterized as (see Figs. 1 and 2),
\begin{figure}
\includegraphics[width=9cm]{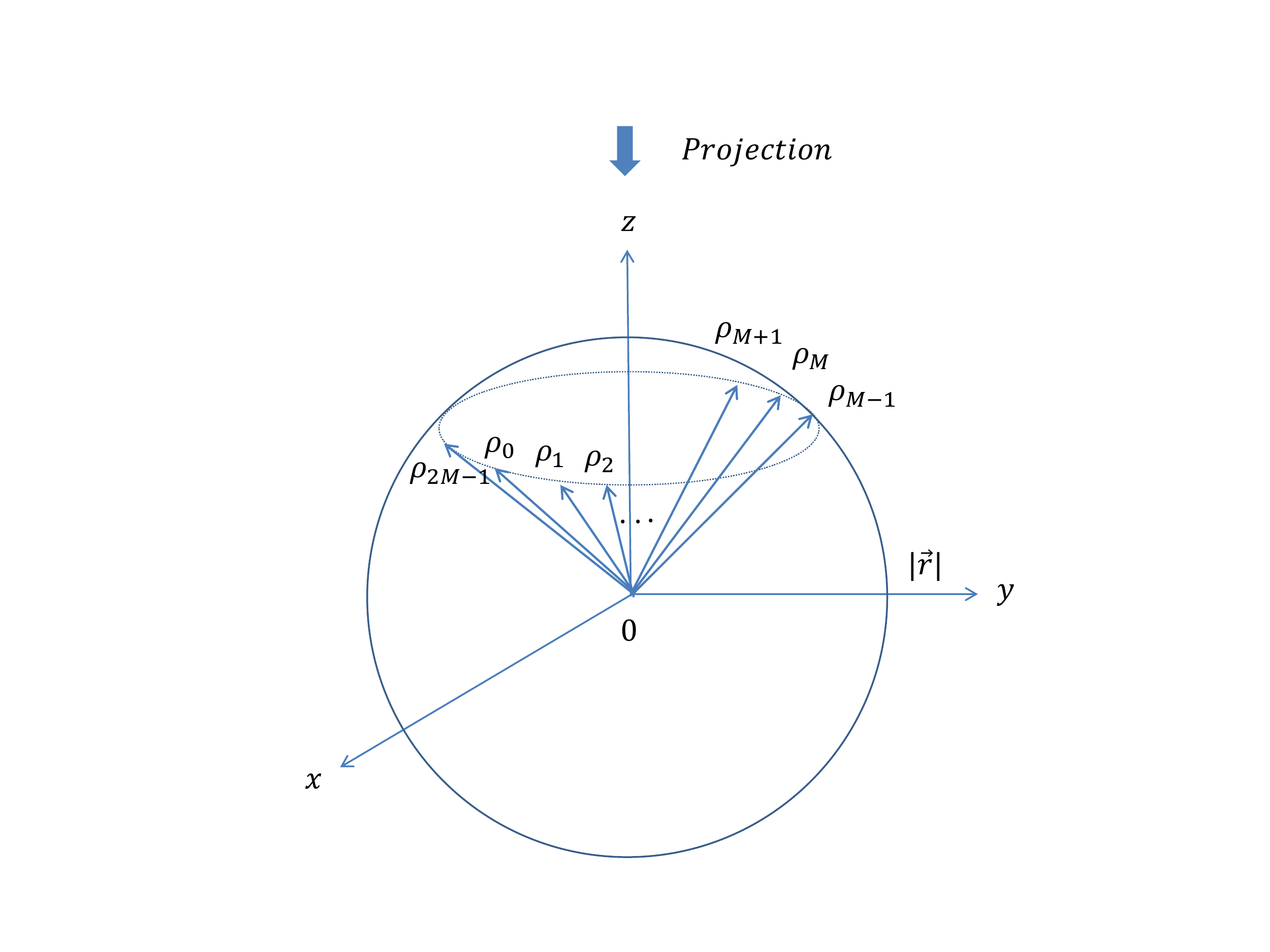}
\caption{Representation of the states to be discriminated on the Bloch sphere. $\rho_i$ denotes $\vec{r}(\rho_i)$ which is Bloch vector of qubits to be discriminated. $\rho_i$'s are symmetrically oriented.}
\label{Fig-1}
\end{figure}
\begin{figure}
\includegraphics[width=9cm]{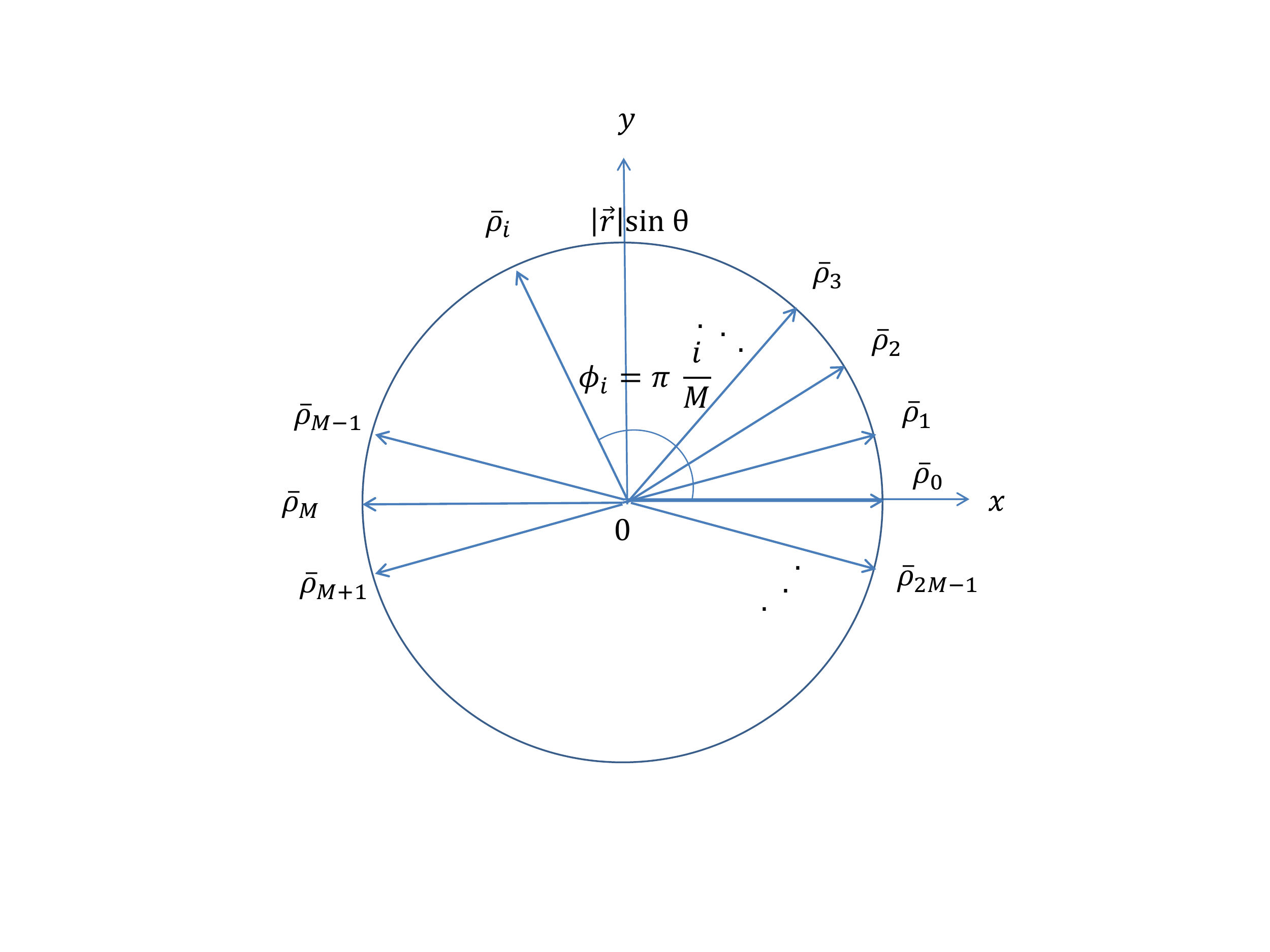}
\caption{Projection of Bloch sphere of Fig. 1 onto the xy plane. $\rho_i$ is projected on xy plane to be $\bar{\rho}_i$.}
\label{Fig-2}
\end{figure}
\begin{equation}
\rho_i=  \frac{1}{2} \{\openone +\vec{r}(\rho_i) \cdot \vec{\sigma}\}, \hspace{5mm} i=0,1,2,..., 2M-1.
\label{B}
\end{equation}
Here
\begin{equation}
\vec{r}(\rho_i)= |\vec{r}| (\sin \theta \cos \phi_i, \sin \theta \sin \phi_i, \cos \theta),
\label{C}
\end{equation}
where $\phi_i=\pi (i/M)$. We consider only the symmetric case where $q_i$ are all equal, and $q_i= 1/(2M)$.

Let us consider a communication scenario in which QSD is incorporated \cite{Her82,Gis98,Han10}.

{\it Proposition 1}. If two different decompositions of states corresponding to the same density operator can be discriminated, superluminal communication can be achieved between two parties sharing appropriate entangled states:
Assume that Alice and Bob are sharing an ensemble of entangled states. In this case Bob's reduced density operator $\rho_B$ is fixed. Consider two different decompositions of states $\{p_k, |\phi^k \rangle \langle \phi^k|\}$ and $\{q_l, |\psi^l \rangle \langle \psi^l|\}$ corresponding to the $\rho_B$. (Here $\sum_k p_k=\sum_l q_l=1$ and $ p_k, q_l>0$.) That is, $\sum_k p_k |\phi^k \rangle \langle \phi^k| = \sum_l q_l |\psi^l \rangle \langle \psi^l|=\rho_B$.
By  the Gisin-Hughston-Jozsa-Wootters theorem \cite{Gis89, Hug93}, however, Alice can generate any decomposition she wants by performing an appropriate measurement. That is, there exists a measurement $M_0$ (the other measurement $M_1$) such that if Alice performs $M_0$ ($M_1$) on her quantum states then a decomposition $\{p_k, |\phi^k \rangle \langle \phi^k|\}$ ($\{q_l, |\psi^l \rangle \langle \psi^l|\}$) is generated at Bob's site. They can communicate as follows. If Alice wants to send a bit $0$ ($1$), she performs $M_0$ ($M_1$). $\Box$

Now let us consider a decomposition, $(1/2) \rho_m + (1/2) \rho_{2M-m} $ that corresponds to a density operator $\frac{1}{2} (\openone +\vec{r}_0 \cdot \vec{\sigma})$. Here $\vec{r}_0= \frac{1}{2} \vec{r}(\rho_m)+ \frac{1}{2} \vec{r}(\rho_{2M-m})$ and $m=1,2,...,M-1$ (see Figs. 1 and 2).
By a simple geometric argument, we can find a $p>0$ such that
\begin{equation}
 \frac{1}{2} \vec{r}(\rho_m)+ \frac{1}{2} \vec{r}(\rho_{2M-m})= p \vec{r}(\rho_0)+ (1-p) \vec{r}(\rho_M).
\label{D}
\end{equation}
An angle between $\rho_m$ and $\rho_0$ in the projected plane is $\phi_m= \pi (m/M)$. Now we get
\begin{equation}
p=\frac{\cos\phi_m +1}{2}=\cos^2 \frac{\phi_m}{2}.
\label{D-2}
\end{equation}
By Eqs. (\ref{A}) and (\ref{D}), clearly we have
\begin{equation}
\frac{1}{2} \rho_m + \frac{1}{2} \rho_{2M-m}= p \rho_0+ (1-p) \rho_M.
\label{E}
\end{equation}

On the other hand, suppose we have a device for QSD, a ``state discriminator''. Here for a moment we neither specify a figure of merit nor require that the state discriminator is an optimal one. However, we assume that the state discriminator has symmetry, $P(i|j)= P(j|i)$ and $P(i|j)= P(i+n|j+n)$  (mod $2M$). It is not that all state discriminators should be symmetric even if the input states has symmetry as in our case. Consider a (useless) state discriminator that always gives a fixed output $0$ for any input, which is clearly not symmetric. However, as we show below, for any asymmetric state discriminator there exists a symmetric one whose score is the same as the score of the asymmetric one. Hence it is sufficient for us to consider only symmetric one because our goal is to find optimal state discriminator.

Now let us show that, for any asymmetric state discriminator, there exists a symmetric state discriminator whose score is the same as the one by the asymmetric one. Let us denote conditional probabilities of the asymmetric one by  $P(j|i)$, which gives a score
\begin{eqnarray}
S = \sum_{i=1}^{N} \frac{1}{N} \{\sum_{j=1}^{N} f(i,j) P(j|i)\}= \frac{1}{N} \sum_{i,j} f(i,j) P(j|i).
\label{E-2}
\end{eqnarray}
Let us consider a symmetric one with
\begin{eqnarray}
P^{\prime}(j|i) &=& \frac{1}{2N} \{\sum_{n=1}^{N}P(j+n|i+n)
\nonumber\\
 &&+\sum_{m=1}^{N}P(i+m|j+m)\} \hspace{2mm} (\mbox{mod} N).
\label{E-3}
\end{eqnarray}
We can see that score by $P^{\prime}(j|i)$ is the same as score by $P(j|i)$:
\begin{eqnarray}
&& \frac{1}{N} \sum_{i,j} f(i,j) P^{\prime}(j|i) \nonumber\\
&=& \frac{1}{N} \sum_{i,j} f(i,j) \frac{1}{2N} \{\sum_{n=1}^{N}P(j+n|i+n)
\nonumber\\
&& \hspace{25mm}+\sum_{m=1}^{N}P(i+m|j+m)\}
\nonumber\\
&=& \frac{1}{2N^2} \sum_{i,j} \{ \sum_{n=1}^{N} f(i+n,j+n) P(j+n|i+n)
\nonumber\\ && \hspace{12mm}+\sum_{m=1}^{N} f(i+m|j+m) P(i+m|j+m)\}
\nonumber\\
&=& \frac{1}{2N^2} \{ 2N \sum_{i,j} f(i,j) P(j|i) \}
\nonumber\\
&=& \frac{1}{N} \sum_{i,j} f(i,j) P(j|i)
\label{E-4}
\end{eqnarray}
Here symmetry of the figure of merits is used to get second equality.

The state discriminator we consider is a ``black box,'' which can include quantum measurement device and anything else helpful for the task. For a certain input state $\rho_i$, the state discriminator will give an output $j$ as the optimal guess. 

Now let us consider the left-hand and right-hand sides of Eq. (\ref{E}) as the two different decompositions in the context of Proposition 1. By the no-signaling principle and Proposition 1, the two decompositions cannot be discriminated by any means. Therefore the two decompositions cannot be discriminated by the state discriminator. For an output 0, this implies
\begin{equation}
\frac{1}{2} P(0|m) + \frac{1}{2} P(0|2M-m)= p P(0|0)+ (1-p) P(0|M)
\label{F}
\end{equation}
by linearity of quantum mechanics. Here $P(0|0)$ and $P(0|M)$ are constants and thus they can be set to be $A$ and $B$, respectively. By symmetry, we have $P(0|m)= P(0|2M-m)$. From Eqs. (\ref{D-2}) and (\ref{F}),
\begin{equation}
 P(0|m)=  A \cos^2 \frac{\phi_m}{2} + B \sin^2 \frac{\phi_m}{2}.
\label{G}
\end{equation}
By setting $A-B= \alpha$ and $B=\beta$, we have
$P(0|m)= \alpha \cos^2 (\phi_m/2) + \beta= \alpha \cos^2 \{(m \pi)/2M\}+ \beta$.
By symmetry, we have, for each $i$ and $j$,
\begin{equation}
P(i|j)= P(j|i)= \alpha \cos^2 \frac{|i-j|\pi}{2M}+ \beta.
\label{H}
\end{equation}
Note that a state discriminator with \{$A=\gamma, B=\delta$\} and another with \{$A=\delta, B=\gamma$\} can be interconverted to each other by simply re-labeling the output, $j \rightarrow j+M$, so they are equivalent to each other. Therefore, we confine ourselves to the case $A-B>0$ without loss of generality.
We can see that as $|A-B|/\mbox{min} (A,B)$ becomes larger, the function becomes narrower.  Here $\mbox{min} (A,B)$ is the minimal one between $A$ and $B$. The case when $B=0$ and $A> 0$ gives maximal information.
Conditional probability of this case also gives maximal score.
In the case of pure states, the optimal conditional probability is achieved by the uniform measurement, which thus becomes an optimal measurement \cite{Han10}. In our case, however, we can expect that the conditional probability would have a more broad form because the states to be discriminated are mixed states and thus more nonorthogonal.
\subsection{Bounding conditional probability and obtaining optimal measurement}
Before we discuss how to get the more broad one, let us refresh QSD by the no-signaling principle. Here what we need to do first is to construct a set of states $\sigma_i$ such that
\begin{equation}
p_i \rho_i + (1-p_i) \sigma_i= p_j \rho_j + (1-p_j) \sigma_j \equiv L
\label{I}
\end{equation}
for all $i,j$'s and
\begin{equation}
q_i= \frac{p_i}{\sum_i p_i}.
\label{K}
\end{equation}
{\it Proposition-2} \cite{Bae11}. From the no-signaling principle, the guessing probability in any state discriminator must be bounded as
\begin{equation}
P_{\mbox{guess}}= \sum_{i} q_i P(i|i) \leq \frac{1}{\sum_{i} p_i}.
\label{L}
\end{equation}
Assume that Alice and Bob are sharing an ensemble of an entangled state for which $\rho_B= L$. By the theorem of Gisin-Hughston-Jozsa-Wootters \cite{Gis89, Hug93}, Alice can generate any decomposition, $p_i \rho_i + (1-p_i) \sigma_i \equiv L_i$ that she wants at Bob's site. Now let $P_D(j|i)$ be the conditional probability that an output $j$ is given by a state discriminator when $L_i$ is generated. Note that the state discriminator gives a certain output even when $\sigma_i$ is input. Hence we have $p_i P(i|i) \leq P_D (i|i) $. By the no-signaling principle, we have $P_D(j|i)=P_D(j|i')$ for all $i,i',j$. Here $i'=0,1,...,N$. Thus we have $\sum_i P_D(i|i)= \sum_i P_D (i|1)=1$. By combining Eq. (\ref{K}) and the above equations, we have $P_{\mbox{guess}}= \sum_{i} q_i P(i|i) = \sum_{i} \{p_i/(\sum_i p_i)\} P(i|i) \leq
\{1/(\sum_i p_i) \} \sum_{i} P_D(i|i) = 1/(\sum_i p_i)$. $\Box$

Now let us derive bounds on the conditional probability by using Proposition 2. In our case in which all $q_i$'s are the same, all $p_i$'s are the same by Eq. (\ref{K}). The bound reduces to $P_{\mbox{guess}} \leq 1/(Np)$ when $p_i \equiv p$ \cite{Hwa10}. The $p$ in our case is calculated as
$p= 1/(1+ |\vec{r}||\sin \theta|)$ \cite{Hwa10}.
Thus the bound is
\begin{equation}
P_{\mbox{guess}} \leq \frac{1}{2M} (1+ |\vec{r}||\sin \theta|).
\label{N}
\end{equation}
Note that Eq. (\ref{N}) is valid for any state discriminator.
One may say that Eq. (\ref{N}) is bound for normal QSD. However, the state discriminator made for any figure of merit can also work  for normal QSD.

Now we can see that by some calculations including normalization for $\alpha$ and $\beta$, if the function in Eq. (\ref{H}) becomes narrower than
\begin{equation}
Q(|i-j|) \equiv \frac{|\vec{r}||\sin \theta|}{M} \cos^2 \frac{|i-j|\pi}{2M}+ \frac{1}{2M}(1-|\vec{r}||\sin \theta|),
\label{O}
\end{equation}
then the function violates the bound in Eq. (\ref{N}).
Hence the conditional probability $P(i|j)$ cannot be narrower than $Q(|i-j|)$. Namely, possible ranges are $0 \leq \alpha \leq (|\vec{r}||\sin \theta|)/M$ and $(1- |\vec{r}||\sin \theta|)/(2M) \leq \beta \leq 1/(2M)$. Thus $Q(|i-j|)$ is the optimal conditional probability in our case.
All results so far are valid for any figure of merit since we have not specified a figure of merit yet. Remarkably, conditional probability has the same form in Eq (\ref{H}) for any figure of merit.

Now let us consider a specific figure of merit. We can say that almost (useful) figure of merit should have monotonicity: The figure of merit increases with decreasing $\phi_i$. Provided that the monotonicity is satisfied, the function $Q(|i-j|)$ gives the highest average score. Now if we find a measurement that realize the function $Q(|i-j|)$, the measurement becomes an optimal one.

Now let us discuss how the function $Q(|i-j|)$ is achieved. Let us try the optimal measurement in the normal QSD for our states. The optimal measurement is a symmetric one whose  Positive operator-valued measure (POVM, \cite{Nie00}) is
\begin{equation}
\hat{E}_i = \frac{1}{2M} (\openone +\hat{\delta}_i \cdot \vec{\sigma}),
\label{P}
\end{equation}
where $\hat{\delta}_i= \{\cos(\pi i/M), \sin(\pi i/M), 0\}$ is a unit vector in the $xy$ plane. After direct calculation, we can see that the measurement gives the function $Q(|i-j|)$. Hence the measurement by Eq. (\ref{P}) is an optimal measurement for a state discriminator with monotonous figure of merits, and the function $Q(|i-j|)$ is the corresponding optimal conditional probability.
\section{Discussion and Conclusion}
The result applies to any (symmetric) black box which gives an outcome $j$ for the input states $\rho_i$ with fixed conditional probability $P(j|i)$, regardless of whether the black-box is made for state discrimination. That is, conditional probability of any such black box must have the form in Eq. (\ref{H}).

In some cases like in the case of information gain \cite{Tar99}, the score cannot be written in the form of Eq. (\ref{B}). However, the result here is applicable. As discussed above, conditional probability has the form in Eq. (\ref{H}). A conditional probability within allowed range of $\alpha$, which maximize the score, is the optimal one. In the case of information gain the optimal one should be one with maximal $\alpha$.

We considered QSD with general (symmetric) figure of merit. We first solved the problem for an even number of symmetric qubits with use of the no-signaling principle. Here both methods used for QSD \cite{Hwa05, Bae08, Hwa10, Bae11} and QSE \cite{Han10} are combined to get the solutions. We showed that, with use of the no-signaling principle, conditional probability always has the form $\alpha \cos^2 (\phi/2)+ \beta$. We tightened the range for the conditional probability with use of the no-signaling principle again. Remarkably, these results are valid for any (symmetric) figure of merit.
The optimal conditional probability $Q(|i-j|)$ is achieved by a simple symmetric measurement. Therefore, the measurement is the optimal one for our case.

\section*{Acknowledgement}
This study was supported by Basic Science Research Program through the National Research Foundation of Korea (NRF) funded by the Ministry of Education, Science and Technology (2010-0007208). This study was financially supported by Woosuk University.

\end{document}